\documentstyle[psfig]{europhys}

\def\And{{\rm and\ }}

\def\stars{\bigskip\centerline{***}\medskip}

\newif\ifboo \boofalse

\def\Review#1{\boofalse{\it #1},}
\def\Name#1{{\sc #1},}
\def\Vol#1{\ifboo Vol. {\bf #1}\else{\bf #1}\fi}
\def\Year#1{\ifboo #1\else(#1)\fi}
\def\Book#1{\bootrue{\it #1},}
\def\Page#1{\ifboo {\rm p. #1}\else{\rm #1}\fi}

\begin{document}

\euro{}{}{}{1998}
\Date{}
\shorttitle{N.~V.~Brilliantov and T.~P\"oschel ROLLING FRICTION 
 OF A VISCOUS SPHERE ON A HARD PLANE}

\title{Rolling friction of a viscous sphere on a hard plane}

\author{Nikolai V. Brilliantov\inst{1,2}\footnote{email 
nbrillia@summa.physik.hu-berlin.de} \And Thorsten 
P\"oschel\inst{2}\footnote{email 
thorsten@itp02.physik.hu-berlin.de}}

\institute{
\inst{1} Moscow State University, Physics Department, 
   Moscow 119899, Russia\\
\inst{2} Humboldt-Universit\"at zu Berlin, Institut f\"ur Physik, 
   Invalidenstra\ss e 110,\\ D-10115 Berlin, Germany
}

\rec{}{}

\pacs{
\Pacs{46}{30.Pa}{Friction, wear, adherence, hardness, mechanical 
contacts, tribology}
\Pacs{62}{40.+i}{Anelasticity, internal friction, stress 
relaxation, and mechanical resonances}
\Pacs{81}{40.Pq}{Friction, lubrication, and wear}
}

\maketitle
\begin{abstract}
  A first-principle continuum-mechanics expression for the rolling
  friction coefficient is obtained for the rolling motion of a
  viscoelastic sphere on a hard plane. It relates the friction
  coefficient to the viscous and elastic constants of the sphere
  material. The relation obtained refers to the case when the
  deformation of the sphere $\xi$ is small, the velocity of the
  sphere $V$ is much less than the speed of sound in the material
  and when the characteristic time $\xi/V$ is much larger than the
  dissipative relaxation times of the viscoelastic material. To our
  knowledge this is the first ``first-principle'' expression of the
  rolling friction coefficient which does not contain empirical
  parameters.
\end{abstract}

Rolling friction is one of the basic phenomenon men encounters in his
everyday life since antediluvian times when the wheel was invented.
The phenomenon of rolling friction has been interesting to scientist
for a long time. Scientific publications on this subject range back to
(at least) 1785 when Vince described systematic experiments to
determine the nature of friction laws~\cite{Vince:1785}, and important
scientists dealt with this problem among them
O.~Reynolds~\cite{Reynolds:1874}.  The rolling friction is of great
importance in engineering and science. From the speed of landslides
Huang and Wang~\cite{HuangWang:1988} argued that the rolling friction
plays an important r\^{o}le even in geological processes, for a
theoretical consideration see~\cite{HerrmannManticaBessis:1990}. For
its major importance the phenomenon has been studied intensively by
engineers and physicists (e.g.~\cite{Engineers}), however,
surprisingly few is known about its basic mechanisms. To our knowledge
there is still no ``first-principle'' expression for the rolling
friction coefficient available which relates this coefficient only to
the material constants of the rolling body and does not contain
empirical parameters.

It has been shown that surface effects like adhesion~\cite{adhesion},
electrostatic interaction~\cite{DeryaguinSmilga:1994}, and other
surface properties~\cite{surface} might influence the
value of the rolling friction coefficient. Theoretically this problem
was studied in Ref.~\cite{FullerRoberts:1981} where the authors
propose a model of a surface with asperities to mimic friction
(see also~\cite{PoeschelHerrmann:1993}). In other
studies~\cite{greenwood,bulkRolling} it was argued that for 
viscoelastic materials ``the rolling friction is due very
little to surface interactions:~the major part is due to deformation
losses within the bulk of the material''~\cite{greenwood}. Based on
this concept the rolling friction coefficient was calculated
in~\cite{greenwood} where the deformation in the bulk was assumed to be
completely plastic; then an empirical coefficient was introduced to
account for the retarded recover of the material.

In the present letter we also consider the rolling friction as a
phenomenon appearing due to viscous processes in the bulk. We assume
that energy losses due to surface effects may be neglected, compared
to the viscous dissipation in the bulk. Thus we attribute the effect
of rolling friction to viscous dissipation in the material due to
time-dependent deformation. We assume that the only r\^{o}le of
surface forces is to keep the rolling body from sliding. We use a
quasi-static approach~\cite{BrilliantovSpahnHertzschPoeschel:1994} and
obtain the rolling friction coefficient.
\begin{figure}[htbp]
  \centerline{\psfig{file=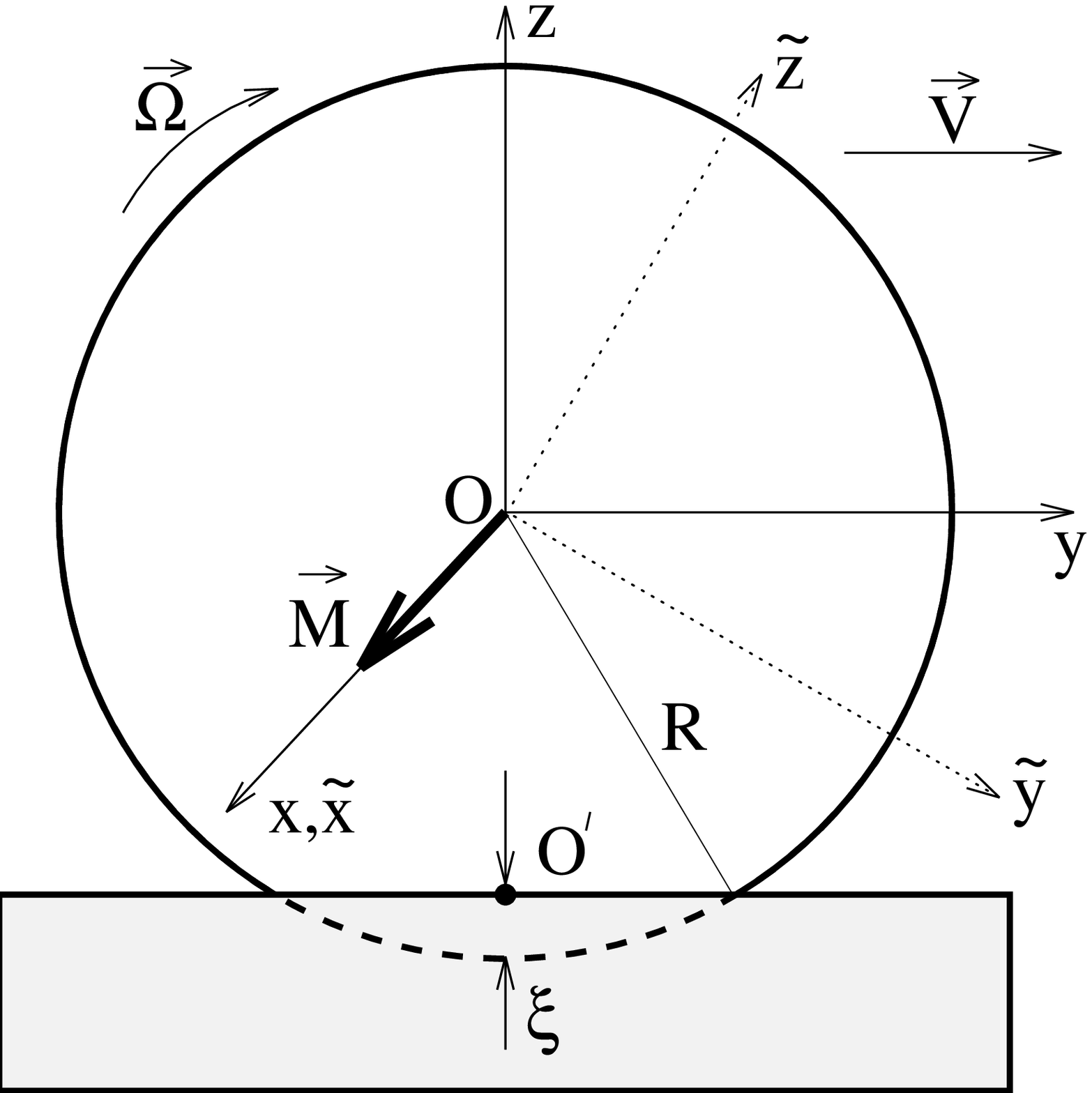,width=5cm}\hspace{1cm}
    \psfig{file=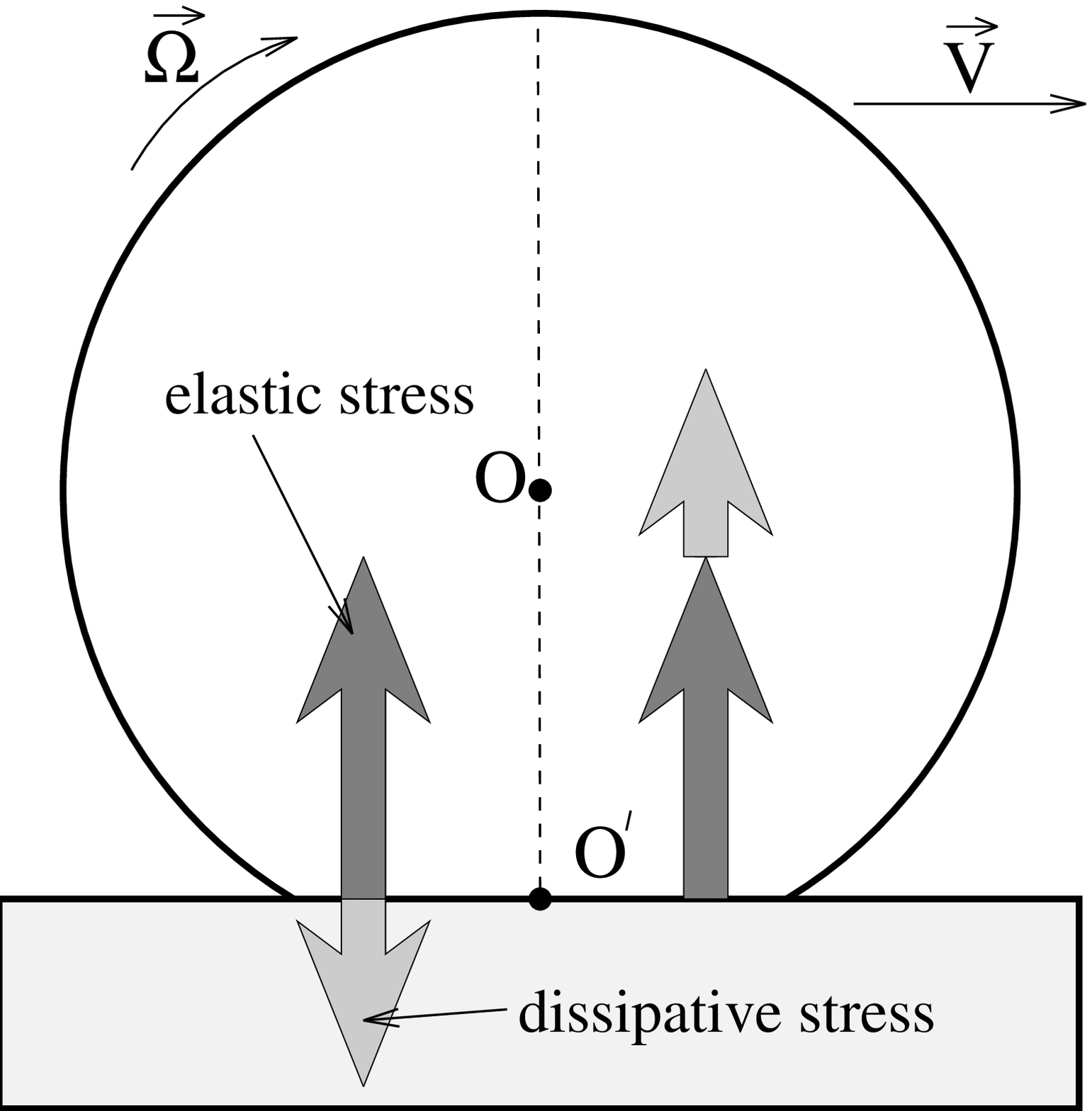,width=5cm}}
  \caption{Sketch of the rolling sphere, the right figure shows the 
    appearance of the friction momentum: in the front the elastic
    stress is enhanced by the dissipative stress, while in the rear it
    is diminished.}
  \label{fig:sketch}
\end{figure}

Fig.~\ref{fig:sketch} shows a sketch of the sphere of radius $R$ and
mass $m$ rolling with angular velocity ${ \Omega}$ which is related
to its linear velocity ${ V}$ as $ V=\Omega R$. The center of the
coordinate system $OXYZ$ coincides with the center of the sphere
(moving coordinate system), so that the plane is normal to the
$Z$-axis and located at $z_p <0 $. Within our model we assume that the
surface is much harder than the sphere, hence, the surface remains flat
in the contact area while the sphere is deformed. Thus the contact
area is a circle of radius $a$. The compression $\xi=R-|z_p|$ of the
sphere causes the displacement field in the bulk of the material
${\vec{u}} ( {\vec{r}}, t )$ where ${\vec r}$ gives the location of
the point within the sphere at time $t$. The displacement field gives
rise to a (tensorial) field of deformation $\vec{\nabla} \circ
\vec{u}$, and thus to the elastic part of the stress tensor
$\hat{\sigma}_{el}$~\cite{LandauElstitcity:1965}
\begin{equation}
\hat{\sigma}_{el}= E_1~ \left[\frac{1}{2}
                    \left\{ \vec{\nabla} \circ \vec{u} + 
                    \vec{u} \circ \vec{\nabla} \right\} -
                    \frac{1}{3} \hat{I} \, \vec{\nabla} \cdot 
                    \vec{ u} \right] + 
E_2\,\hat{I} \, \vec{ \nabla } \cdot \vec{ u}\,.
\label{elast}
\end{equation}
Here, $\hat{I}$ is the unit tensor and $E_1=\frac{Y}{1+\nu }$ and
$E_2=\frac{Y}{3(1-2\nu)}$ denote the elastic material constants with
$Y$ and $\nu$ being the Young modulus and the Poisson ratio,
respectively. Since the deformation depends on time via the
time-dependent field $\vec{ u} ( \vec{ r}, t )$, the dissipative part
of the stress tensor reads
\begin{equation}
\hat{\sigma}_{dis}= \eta_1~ \left[\frac{1}{2}
                    \left\{\vec{ \nabla} \circ \dot{ \vec{ u}} + 
                    \dot{\vec{ u}} \circ \vec{ \nabla } \right\} -
                    \frac{1}{3} \hat{I}\,\vec{ \nabla} 
                    \cdot \dot{\vec{ u}} \right] +
\eta_2\,\hat{I}\, \vec{ \nabla}  \cdot \dot{\vec{ u}}\,,
\label{diss}
\end{equation}
where $\dot{ \vec{ u}}$ denotes the time derivative of the
displacement field and $\eta_1$ and $\eta_2$ are the viscous
constants of the sphere material~\cite{LandauElstitcity:1965}. The
total stress $\hat{\sigma}$ is a sum of both parts,
$\hat{\sigma}=\hat{\sigma}_{el}+\hat{\sigma}_{dis}$. The force
normal to the plane which acts on the sphere is caused by its own weight
$mg$. It reads
\begin{equation}
\vec{ F}= \int_{S_a} \hat{\sigma} \cdot \vec{ n}\, ds = \vec{ n} 
\int \int dx dy \, \left(\sigma_{el}^{\mbox{\footnotesize\em zz}}+ 
\sigma_{\mbox{\footnotesize\em dis}}^{\mbox{\footnotesize\em zz}} \right) 
=\vec{ n} F^{N}\,,
\label{force}
\end{equation}
where $\vec{ n}$ is the (positively directed) unit normal to the
plane, $\sigma_{el}^{\mbox{\footnotesize\em zz}}$ and 
$\sigma_{\mbox{\footnotesize\em dis}}^{\mbox{\footnotesize\em zz}}$ are 
the corresponding components of the stress tensors. The integration is to be
performed over the contact circle $S_a$ on the plane $z=z_p$.
Correspondingly, the torque acting on the sphere with respect to the
point $O'$ at rest (no slipping conditions) is given by
\begin{equation}
\vec{ M}= \int_{S_a} \left( \vec{ r}- \vec{ r}_{O'} \right)
 \times \left( \hat{\sigma} \cdot \vec{ n} \right)  ds \,,
\label{torque}
\end{equation}
where $ \vec{ r}_{O'}= \left( 0,0,z_p \right)$. The  torque $\vec{ M}$ 
thus characterizes the rolling friction. 

Let the sphere move in $y$-direction, $\vec{ V}= \left(0,V,0 \right)$,
then the angular velocity and the friction torque are directed along
the $x$-axis, i.e. $\vec{ \Omega}= \left( -\Omega, 0,0 \right)$ and
$\vec{ M}=\left( M, 0,0 \right)$ (see Fig.~\ref{fig:sketch}). For this
geometry from Eq.~(\ref{torque}) follows
\begin{equation}
 M= \int\int dx dy\,\, y  \,\hat{\sigma}^{\mbox{\footnotesize\em zz}} 
\left( x,y,z=z_p \right) 
\label{torque1}
\end{equation}
The integration in Eq.~(\ref{torque1}) again is performed over the
contact circle.

To perform the calculation of the friction torque one needs the
displacement field $\vec{ u} ( \vec{ r}, t )$ and the field of the
displacement velocity $\dot{\vec{ u}} ( \vec{ r}, t )$. Generally, to
find these quantities is a rather complicated problem. Nevertheless,
one can use a quasi-static approximation, provided that the
displacement velocities in the bulk are much smaller than the speed of
sound in the sphere material and provided the characteristic time of
the process of interest, estimated by $\tau=\xi/V$, is much larger
than the dissipative relaxation times of the material
(see~\cite{BrilliantovSpahnHertzschPoeschel:1994} for details). 
This allows to employ for the displacement field $\vec{ u} $ 
the results of the static problem, i.e. of the 
Hertz contact problem which refers to a ``slow'' elastic collision 
of two spheres~\cite{Hertz:1882}. 
In the case considered here one of the
spheres (the plane) should have infinite radius.  
Let $\vec{u}_{el}(\vec{ r}) $ be the solution of this (static) contact
problem. It corresponds to the displacement field in a sphere under
the compression $\xi$ (see Fig.~\ref{fig:sketch}) if it would be at
rest.  In the quasi-static approximation the distribution $\vec{
  u}_{el}(\vec{ r}) $ persists (i.e. it does not change with time) in
the coordinate system $OXYZ$, moving with the velocity $V$ (with no
rotation).  In a body-fixed coordinate system $O
\tilde{X}\tilde{Y}\tilde{Z}$ (see Fig.~\ref{fig:sketch}), which
rotates with the angular velocity $\Omega$ and coincides at some time
instant with $OXYZ$ the displacements distribution is the same as in
$OXYZ$ ~(~i.e. $\vec{ u}_{el}(\vec{ r}) $~), while the displacement
velocity distribution follows from the kinematic equation 
\begin{equation}
\dot{\vec{ u}} ( \vec{ r} )=( \vec{ \Omega} \cdot \vec{ r} 
\times \vec{ \nabla} ) 
\vec{ u}_{el}(\vec{ r})=-\Omega \left( y\,\partial_z-z\, \partial_y 
\right)\, \vec{ u}_{el}(\vec{ r}) \,
\label{kinematic}
\end{equation}
which relates both coordinate systems;
here  $\partial_x=\frac{\partial}{\partial x}$,
$\partial_y=\frac{\partial}{\partial y}$ and
$\partial_z=\frac{\partial}{\partial z}$. The elastic part of the
stress tensor is then obtained by substituting $\vec{ u}_{el}(\vec{
  r}) $ into Eq.~(\ref{elast}). This part of the stress tensor 
corresponds to the static case when no torque acts on the sphere. 
Therefore, in the quasi-static approximation only the dissipative 
part of the stress
tensor $\hat{\sigma}^{\mbox{\footnotesize\em
    zz}}_{\mbox{\footnotesize\em dis}}$ should be used in
Eq.~(\ref{torque1}) for the friction torque.  Using kinematic
Eq.~(\ref{kinematic}) and definitions (\ref{elast}) and (\ref{diss})
one finds for the dissipative part
\begin{eqnarray}
\hat{\sigma}^{\mbox{\footnotesize\em zz}}_{\mbox{\footnotesize\em dis}}&=&
-\Omega \left\{  \left( y \,\partial_z-z \, \partial_y \right)
 \hat{\sigma}^{\mbox{\footnotesize\em zz}}_{el} 
  \left( E_1 \leftrightarrow \eta_1 ; 
  E_2 \leftrightarrow \eta_2 \right) + \right. \nonumber \\
  && \left. + 2\hat{\sigma}^{yz}_{el} 
  ( E_1 \leftrightarrow (\eta_2- \eta_1/3) ; E_2 \leftrightarrow \eta_2 ) 
  - (2\eta_2+ \eta_1/3 )\partial_y  u^{z}_{el}  \right\}. 
\label{stressdiss}
\end{eqnarray}
where $\vec{ u_{el}}= \left( u^x_{el}, u^y_{el}, u^z_{el} \right)$ and
$E_1 \leftrightarrow \eta_1$, $E_2 \leftrightarrow \eta_2$, etc.  in
the arguments of the {\it elastic} stress tensor denote that the
elastic constants $E_1$, $E_2$, etc. in Eq.~(\ref{elast}) should be replaced
by the dissipative constants $\eta_1$, $\eta_2$, etc.

For the case of small compression $\xi/R \ll 1$ which is addressed here we 
estimate the magnitude of different terms in Eq.~(\ref{stressdiss})
taken at the plane $z=z_p \approx -R$, since these values of the 
stress tensor determine the friction torque (see Eq.(\ref{torque1})).
For the diagonal component of the stress tensor one
has the known solution for the Hertz contact
problem~\cite{Hertz:1882,LandauElstitcity:1965}
\begin{eqnarray}
\label{elstressgen}
\sigma^{\mbox{\footnotesize\em zz}}_{el}(x,y,z_p) &=& 
 E_1{\partial_z{u_{el}^z}} + 
\left( E_2-{E_1}/3\right) \left( {\partial_x {u_{el}^x}} +
{\partial_y {u_{el}^y}} + {\partial_z{u_{el}^z}} \right) \nonumber \\
&=& \frac32\frac{F^{N}_{el}}{\pi ab}\sqrt{1-\frac{x^2}{a^2}
-\frac{y^2}{b^2}}\,.
\end{eqnarray}
where $a=b=\sqrt{2\xi R}$ is the radius of the contact
circle and $F^{N}_{el}$ is the total elastic force, acting by the
surface (in normal direction) on the sphere:
\begin{equation}
F^{N}_{el}(\xi) = \frac{2}{3}\frac{Y}{\left( 1-\nu ^2\right) }R^{1/2} \,
         \xi^{3/2}~.
\label{Hertz}
\end{equation} 
From Eq.~(\ref{Hertz}) follows $\sigma^{\mbox{\footnotesize\em
    zz}}_{el} \sim \left(\xi/R \right)^{1/2}$ at $z=z_p$.  Applying
geometric consideration of the deformation of the sphere one gets
$u^x_{el} \sim u^y_{el} \sim \xi^{3/2}/R^{1/2}$, and $u^z_{el} \sim
\xi$. The estimates $\partial_x \sim \partial_y \sim 1/a \sim (\xi
R)^{-1/2}$ and $\partial_z \sim (\xi R)^{-1/2}$ may be obtained too.
The expression for $\partial_z$ follows from the estimates for
$u^z_{el}$, $\sigma^{\mbox{\footnotesize\em zz}}_{el}$ and the
definition of $\sigma^{\mbox{\footnotesize\em zz}}_{el}$. Finally we
find for the deformation
\begin{eqnarray}
\label{deform}
\partial_x u^x_{el} \sim \partial_y u^x_{el} \sim \partial_z u^x_{el} \sim 
\xi/R \ll \left(\xi /R \right)^{1/2} \nonumber \\
\partial_x u^y_{el}  \sim \partial_y u^y_{el} \sim \partial_z u^y_{el} \sim 
\xi/R \ll \left(\xi /R \right)^{1/2} \nonumber \\
\partial_x u^z_{el} \sim  \partial_y u^z_{el} \sim \partial_z u^z_{el} \sim 
 \left(\xi /R \right)^{1/2} \,.
\end{eqnarray}
Eqs.~(\ref{deform}) further yield $\sigma^{yz}_{el} \sim \left(\xi/R
\right)^{1/2}$. Using these estimates one can calculate the
relative magnitude of the terms in Eq.~(\ref{stressdiss}) and
then compare their contribution to the rolling friction torque  $M$.
After these calculations one arrives at
\begin{equation}
\label{M1}
M=-\Omega R \int \int dx dy \,\, y \partial_y \,
\hat{\sigma}^{\mbox{\footnotesize\em zz}}_{el}
\left( E_1 \leftrightarrow \eta_1 ; E_2 \leftrightarrow \eta_2 \right)
\end{equation}
where all other terms which are by a factor $\left(\xi/R
\right)^{1/2}$ smaller are omitted.  Integrating by parts the
right-hand side of Eq.~(\ref{M1}) and taking into account that the
stress tensor vanishes on the boundary of the contact circle one
finally finds
\begin{equation}
\label{M2}
M=\Omega R \left\{ \int \int dx dy  \,\, 
{\sigma}^{\mbox{\footnotesize\em zz}}_{el}
\left( E_1 \leftrightarrow \eta_1 ; 
E_2 \leftrightarrow \eta_2 \right) \right\}
\end{equation}
The expression in the curly brackets in Eq.~(\ref{M2}) would coincide
with the normal elastic force $F^{N}_{el}$ (see Eq.~(\ref{force})) if
it would contain the elastic constants $E_1$ and $E_2$ instead of the
dissipative constants $\eta_1$ and $\eta_2$. To perform the
integration in Eq.~(\ref{M2}) we apply the rescaling procedure
proposed in~\cite{BrilliantovSpahnHertzschPoeschel:1994}. The
coordinates are transformed due to $x= \alpha x'$, $y= \alpha y'$ and
$z=z'$ with
\begin{equation}
\alpha =  \left(\frac{\eta_2-\frac 13\eta_1}
         {\eta_2+\frac 23\eta_1}\right) 
         \left(\frac{E_2+\frac 23E_1}{E_2-\frac 13E_1}\right)\,,
\end{equation}
so that the radius of the contact area in new coordinates is $a=b=
\alpha a'= \alpha b'$. One obtains
\begin{eqnarray}
\label{S1}
{\sigma}^{\mbox{\footnotesize\em zz}}_{el}
\left( E_1 \leftrightarrow \eta_1 ; E_2 \leftrightarrow \eta_2 \right) 
&=& \beta\left( \eta_1\frac{\partial{u_{el}^z}}{\partial z}+
   \left(\eta_2-\frac{\eta_1}3\right)\!\! \left( 
   \frac{\partial {u_{el}^x}}{\partial x} +
   \frac{\partial {u_{el}^y}}{\partial y} +
   \frac{\partial {u_{el}^z}}{\partial z}\right)\! \right) 
   \nonumber\\
&=& \beta \left( E_1\frac{\partial{u_{el}^z}}{\partial z^{\prime}}+
   \left(E_2-\frac{E_1}3\right)\!\! \left( 
   \frac{\partial {u_{el}^x}}{\partial x^{\prime }} +
   \frac{\partial {u_{el}^y}}{\partial y^{\prime }} +
   \frac{\partial {u_{el}^z}}{\partial z^{\prime }}\right)\! \right) 
   \nonumber\\
&=& \beta \frac{3}{2}\frac{F^{N}_{el}}{\pi a^{\prime}b^{\prime}}
   \sqrt{1-\frac{x^{\prime}{}^2}{a^{\prime}{}^2} -
           \frac{y^{\prime}{}^2}{b^{\prime}{}^2}}\nonumber \\
&=&\beta \alpha ^2 \,\frac{3}{2}\frac{F^{N}_{el}}{\pi ab}
   \sqrt{1-\frac{x^2}{a^2}-\frac{y^2}{b^2}}~.
\end{eqnarray}
where $\beta = \frac{1}{\alpha}\frac{\eta_2-\frac 13\eta_1}{E_2-\frac
  13E_1}$.  As it follows from Eq.~(\ref{S1}) after the appropriate
rescaling the desired expression for the stress tensor with the
``replaced'' constants may be written in the same way as in
(\ref{Hertz}). Integration in Eq.~(\ref{M2}) yields the final result
for the torque of the rolling friction in the quasi-static regime
\begin{equation}
M=k_{\mbox{\footnotesize\em rol}}\Omega R F^N  \,.
\label{rol1}
\end{equation}
It may be shown that in the quasi-static regime the normal force $F^N$
which acts on the sphere coincides with $F^{N}_{el}$ of the static
problem.  Fig.~\ref{fig:sketch} explains the appearance the friction
torque in the rolling motion. The rolling friction coefficient
finally reads
\begin{equation}
k_{\mbox{\footnotesize\em rol}}= \frac{1}{3} \, 
\frac{ \left(3 \eta_2-\eta_1\right)^2}
      {\left(3\eta_2+2\eta_1\right)}\, 
      \left[\frac{\left (1-{\nu}^2 \right )\left (1-2 \nu \right)}
      {Y\,\nu^2}\right] \,.
\end{equation}
The latter expression relates the rolling friction coefficient to the
elastic and viscous constants of the rolling body material. 

It is interesting to note that the rolling friction coefficient
$k_{\mbox{\footnotesize\em rol}}$ coincides with the material
constant $A$ in the dissipative force $F_{\mbox{\footnotesize\em
    dis}}^N=\rho \,A \,\sqrt{\xi} \dot{\xi}$,
$\rho=\frac{Y}{(1-\nu^2)}\sqrt{R}$ for two colliding spheres (for
derivation see~\cite{BrilliantovSpahnHertzschPoeschel:1994}).  Using
this expression one can calculate the normal coefficient of
restitution $\epsilon$ for two spheres which collide with relative
velocity $v$~\cite{BrilliantovSpahnHertzschPoeschel:1994}. This
coefficient which measures the energy loss upon collision reads
$1-\epsilon= C_1\,A\,\rho^{2/5}v^{1/5} + \cdots$~
($C_1=1.15344$)~\cite{SchwagerPoeschel:1998}. Therefore, one can find
the rolling friction coefficient from the restitution coefficient of
the same material. To get an estimate of $k_{\mbox{\footnotesize\em
    rol}}$ we take $1-\epsilon=0.1$ for steel particles of mass
$m=10^{-2} \,kg$, colliding with impact velocity $v=1 \, m/s$ to
obtain $\rho^{2/5} = 10^5\,(kg^2/ms^4)^{1/5}$ and
 $k_{\mbox{\footnotesize\em rol}} =
10^{-6}\,s$. Up to our knowledge for the case of a soft sphere rolling
on a hard plane the dependence of the rolling friction coefficient on
the rolling body velocity has not been studied experimentally.

In conclusion, we found for the first time, to our knowledge, the
first-principle expression which relates the rolling friction
coefficient to the elastic and viscous properties of the material. Our
calculations refer to the rolling motion of a viscoelastic sphere on a
hard plane without slipping in the regime when the velocity of the
sphere is much less then the speed of sound in the material and when
the characteristic time of the process $\xi /V$ is much larger than
the dissipative relaxation times of the viscoelastic material.

\stars

The authors want to thank F. Spahn for useful discussion. This work
has been supported by Deutsche Forschungsgemeinschaft (Po 472/3-2).

\end{document}